\documentclass[%
 reprint,
showpacs,
 amsmath,amssymb,
 aps,
]{revtex4-1}

\usepackage[utf8]{inputenc}
\usepackage{graphicx}
\usepackage{dcolumn}
\usepackage{bm}
\usepackage{epstopdf}

\begin{document}

\preprint{APS/123-QED}

\title{Experimental Demonstration of Spectral Intensity Optical Coherence Tomography}

\author{Piotr Ryczkowski$^{1}$}
\author{Jari Turunen$^{2}$}
\author{Ari T. Friberg$^{2}$}
\author{Goëry Genty$^1$}

\affiliation{
$^{1}$Optics Laboratory, Tampere University of Technology, P. O. Box 527, FI-33101 Tampere, Finland \\
$^{2}$Institute of Photonics, University of Eastern Finland, P. O. Box 111, FI-80101 Joensuu, Finland
}

\date{\today}

\begin{abstract}
We demonstrate experimentally quantum-inspired, spectral-domain intensity optical coherence tomography.  We show that the technique allows for both axial resolution improvement and dispersion cancellation compared to conventional optical coherence tomography. The method does not involve scanning and it works with classical light sources and standard photodetectors. The measurements are in excellent agreement with the theoretical predictions. We also propose an approach that enables the elimination of potential artifacts arising from multiple interfaces.

\end{abstract}

\pacs{42.30.Wb; 07.60.Ly; 42.25.Hz; 42.25.Kb}
\maketitle

Optical coherence tomography (OCT) is a powerful, three-dimensional (3D) imaging technique which may be operated in both the spectral and the temporal domain \cite{Leitgeb_TDvFD_OCT, BoerSNR}, and it is widely employed for biological in vitro and in vivo imaging \cite{Fercher1}. The axial resolution of conventional OCT is limited by the spectral bandwidth of the light source and by the dispersion of the optical components and/or the sample under the test. Because of dispersion, especially in fiber-based setups, increasing the spectral bandwidth of the light source may not necessarily lead to resolution improvement \cite{Wojtkowski1}. To circumvent this problem several approaches have been proposed, including both numerical \cite{Fercher2,Banaszek} and experimental techniques \cite{Erkmen,Gouet,Kaltenbeak_QOCTclasic,Mazurek,NasrQOCT2, NasrQOCTdemonstration,NasrQOCT1}. In particular, methods such as quantum-optical coherence tomography (QOCT) or chirped pulse interferometry have been shown to possess built-in dispersion cancelation and resolution enhancement; however, these techniques generally require sophisticated light sources such as single-photon sources or chirped ultrashort pulses and advanced detection techniques \cite{Kaltenbeak_QOCTclasic,Mazurek,NasrQOCT2,NasrQOCTdemonstration,NasrQOCT1}. Further, they operate in the time domain requiring in-depth scanning, which results in slow measurement times.

Intensity-based optical coherence tomography, inspired by QOCT but using any classical broadband light source, was recently put forward in the time domain \cite{Lajunen,Shapiro_conf}. The method produces improved resolution but is hampered by the lack of ultrafast detectors capable of recording rapid intensity variations characteristic of incoherent sources. However, spectral intensity optical coherence tomography (SIOCT) provides a much simpler and cost-effective alternative that operates, without moving parts, in the spectral domain and can use traditional broadband light source of any state of temporal coherence and standard detectors \cite{Ari1,Ari2}. Furthermore, SIOCT requires only a minor modification of the conventional spectral-domain OCT imaging setup while still providing resolution improvement and, more importantly, all even-order dispersion cancellation. Yet, no experimental demonstration of SIOCT has been reported. In this letter, we respond directly to this need and show experimentally the functioning of SIOCT with a classical incoherent source, confirming both the theoretical resolution enhancement and the (group-velocity) dispersion cancellation. The measurements are in excellent agreement with the theoretical predictions. Our results open up new perspectives for high-resolution imaging using classical light sources.

\begin{figure}[b]
\begin{center}
\includegraphics[trim=0.4cm .9cm 0cm .4cm, width=0.5\textwidth]{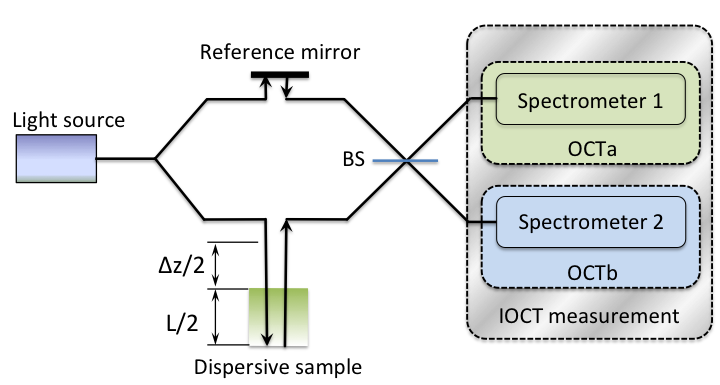}
\end{center}
\caption{(color online) OCT and SIOCT concepts. BS – 50:50 non-polarizing beam splitter. OCT corresponds to a single spectral interference measurement at Spectrometer 1 (or 2). SIOCT corresponds to a simultaneous measurement of spectral interferences at Spectrometers 1 and 2.}
\label{fig:setup1}
\end{figure}

The difference between the principles of the operation of conventional spectral-domain OCT and SIOCT is illustrated in Fig.~\ref{fig:setup1}. Light from an incoherent broadband source is divided between the two arms of an interferometer. One arm serves as the reference while the other contains the sample to be characterized. In traditional OCT, the spectral intensity resulting from the interference of the light in the two arms, recorded at the output of the beam splitter, is of the form
\begin{align}
\label{eq:Ia}
  I_\mathrm{OCTa}(\omega) &\sim |E_\mathrm{S}(\omega) + \mathrm{i}E_\mathrm{R}(\omega)|^{2}, \\
%
\label{eq:Ib}
  I_\mathrm{OCTb}(\omega) &\sim |\mathrm{i}E_\mathrm{S}(\omega) + E_\mathrm{R}(\omega)|^{2},
\end{align}
where $\omega$ is angular frequency, $E_\mathrm{S}$ and $E_\mathrm{R}$ are the complex spectral amplitudes of the electric fields emerging from the sample and reference arms, respectively, and the subscripts a and b denote at which output of the beam splitter the spectral OCT measurement is recorded. The phase difference between the reference and sample arm fields is caused by difference $\Delta z$ in distances that the light propagates in air and the propagation through the dispersive sample of path length $L$. Both distances mark the total path, including wave transits before and after reflection. The intensity of the recorded interferogram can then be written as
\begin{equation}
  I_\mathrm{OCTa}(\omega) = S(\omega)\big| r e^{-\mathrm{i} \left[\omega\Delta z/c +
  \beta(\omega)L \right]} + \mathrm{i} \big|^2,
\label{eq:3}
\end{equation}
where $S(\omega)$ represents the spectrum of the light source, $r=|r|\exp(\mathrm{i}\phi_s)$ is the (complex) amplitude reflection coefficient of the sample, and $c$ the speed of light in vacuum. Further, $\beta(\omega)$ denotes the frequency-dependent propagation constant within the sample. Using a Taylor-series expansion for the propagation constant around the central frequency $\omega _0$ of the source spectrum up to second order, one obtains
\begin{align}
  I_\mathrm{OCTa}&(\omega) = S(\omega_0 + \omega^\prime) \nonumber \\
  &\times \big| r e^{-\mathrm{i} \left[ (\Delta z/c)\omega_0 + \beta_0 L + \tau\omega^\prime
  + (\beta_2L/2)\omega^{\prime 2} \right] } + \mathrm{i} \big|^2,
\label{OCTa_sig}
\end{align}
%
where $\omega = \omega_0 + \omega'$ and $\tau=\Delta z/c+\beta_{1}L$, with $\beta_0$, $\beta_1$, and $\beta_2$ being the propagation constant, the group delay, and the group-velocity dispersion at frequency $\omega_0$, respectively. The envelope of the spectral interference pattern represents the source spectrum whilst the frequency, phase, and amplitude of the modulation underneath depend on the sample position and reflectance, which can then be obtained through a Fourier transform of the recorded interferogram. The resolution is given by the full width at half maximum (FWHM) of the Fourier transform of the term corresponding to the optical path difference between the reference and sample arms. In the absence of dispersion, the axial resolution is inversely proportional to the source bandwidth. With dispersion present in the system, the resolution decreases by a factor $[1 + (\beta_2 L)^2(\Delta\omega/2\sqrt{2\ln 2})^4]^{1/2}$, where $\Delta\omega$ is the FWHM spectral bandwidth of the light source.

In SIOCT, the spectral interference patterns of the fields in the two arms are recorded simultaneously by two separate detectors at the two output ports of the beam splitter, and the location and the reflectance of the sample are obtained from the Fourier transform of the cross product of the individual spectral intensities $C(\omega^\prime) = I_\mathrm{OCTa}(\omega_0 + \omega^\prime) I_\mathrm{OCTb}(\omega_0 - \omega^\prime)$. It is straightforward to show that the interference pattern of the SIOCT signal is then given by
\begin{equation}
  C(\omega^\prime) = S(\omega _0 + \omega^\prime) S(\omega_0 - \omega^\prime)
  [c_0(\omega^\prime) + c_1(\omega^\prime) + c_2(\omega^\prime)],
\label{IOCT_sig}
\end{equation}
where
\begin{align}
  c_0(\omega^\prime) = \:\: &(|r|^2 + 1)^2 \nonumber \\
  &+ 2r^2 \Re \big\{ e^{-\mathrm{i} \left[ 2\omega _0 (\Delta z/c)
  + 2L (\beta _0 - (\beta _2/2) \omega^{\prime 2}) \right]} \big\},
\end{align}
\begin{align}
 c_1(\omega^\prime) = - &4(| r |^2 + 1) \Im \{ e^{-\mathrm{i} \tau \omega^\prime} \} \nonumber \\
 &\times \Re \big\{  e^{-\mathrm{i} \left[ \omega_0 (\Delta z/c)
 + L (\beta _0 + (\beta_2/2) \omega^{\prime 2})\right]} \big\},
\label{eq:c1}
\end{align}
and
\begin{equation}
  c_2(\omega^\prime) = -2|r|^2 \Re \{e^{-\mathrm{i} 2\tau \omega^\prime} \},
\end{equation}
with $\Re$ and $\Im$ denoting the real and the imaginary parts, respectively. One sees that the SIOCT signal produces an interferogram with an envelope bandwidth reduced by a factor of $\sqrt{2}$ compared to that of the conventional OCT inteferogram envelope. The Fourier spectrum of the function $C(\omega^\prime)$ gives access to the sample information and consists of three separated peaks corresponding to the Fourier transforms of $c_0(\omega^\prime)$, $c_1(\omega^\prime)$, and $c_2(\omega^\prime)$ convolved by the Fourier transform of $S(\omega_0+\omega^\prime)S(\omega_0-\omega^\prime)$. Several observations can be made:

We first remark that the product $S(\omega_0+\omega^\prime)S(\omega_0-\omega^\prime)$ is always an even function, even if the spectrum of the light source is not symmetrical with respect to $\omega_0$. The term $c_0(\omega^\prime)$ is an even function whose Fourier transform produces a real-valued peak centered at the zero delay (equal path lengths). The term $c_1(\omega^\prime)$, on the other hand, is an odd function that depends on the optical path length and its Fourier transform corresponds to an imaginary-valued peak centered at the optical path difference between the two arms. Finally, the Fourier transform of $c_2(\omega^\prime)$ gives rise to a real, negative-valued peak at twice the optical path difference. It is precisely this peak in the Fourier spectrum that gives information about the sample position and, after correcting for the distance, one obtains an overall resolution improvement of $\sqrt{2}$ as compared to standard OCT. This also means that the imaging depth is only half of the one obtained by the standard OCT, as the number of data points is constant and their density is doubled in SIOCT. Note further that in practice the terms arising from $c_1(\omega^\prime)$ and $c_2(\omega^\prime)$ can be distinguished as one is real and the other imaginary. Besides the resolution enhancement, another benefit of SIOCT is the inherent dispersion cancellation of all even-order terms in the Taylor-series expansion. This is because in SIOCT intensities of opposite frequencies relative to central frequency $\omega_0$ are multiplied, canceling the even-order phase terms that arise from dispersion.

\begin{figure}[ht]
\begin{center}
\includegraphics[trim=0.4cm .9cm 0cm .4cm, width=0.5\textwidth]{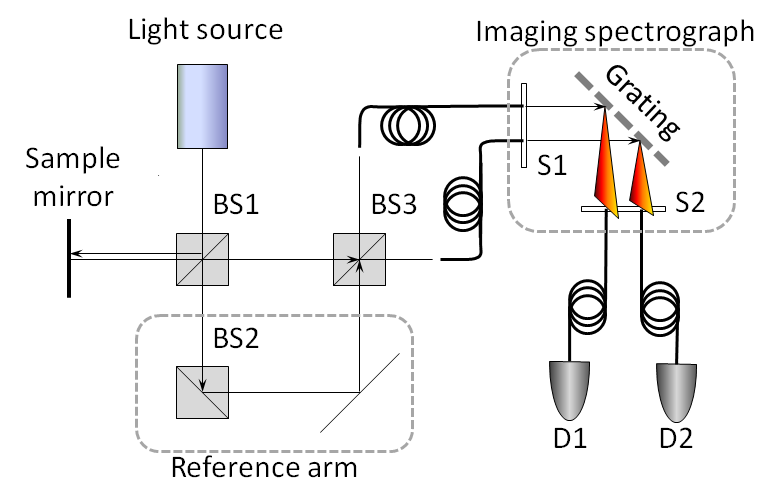}
\end{center}
\caption{(color online) Modified Hong--Ou--Mandel interferometer, experimental setup: BS – 50:50 non-polarizing beam splitter. S1, S2 – spectrograph input and output slits. D1, D2 – photodiodes.}
\label{fig:setup2}
\end{figure}

We confirmed experimentally the resolution improvement and dispersion cancellation as compared to standard OCT using a modified Hong--Ou--Mandel (HOM) interferometer \cite{HOM}, as shown in Fig.~\ref{fig:setup2}. With our setup, the signal recorded at either detector independently is identical to a conventional OCT system, while measuring simultaneously the signal at both detectors allows to construct the SIOCT interferogram. The light source is a fiber-coupled superluminescent diode (Exalos ESL 1620) with a center wavelength of about 1610~nm. The source spectrum is close to Gaussian with a spectral bandwidth of 55~nm (FWHM). The light emitted by the source was collimated with a parabolic mirror and three identical non-polarizing beam-splitter cubes were used to divide the light among the sample and reference arms. A single, partially reflecting mirror placed on a manual translation stage to adjust the optical path length was used as the sample. The reference wave transmitted through beam splitter BS1 was reflected from an additional beam splitter BS2 in order to equalize the dispersion and phase shifts experienced by both beams due to the various optical elements. As a result, any difference between the complex amplitudes of the electric fields in the sample and reference arms before interfering on beam splitter BS3 is caused only by the path difference between the two arms and sample presence. Light at the two output ports of BS3 was coupled to a single-mode fiber using achromatic lenses. The fiber outputs were then placed on top of each other in the object plane of a monochromator (Horiba iHR~550) allowing to measure simultaneously the spectral intensities at the two detectors. On passing through the output slit of the monochromator, light was collected by multimode fibers and intensities were measured by InGaAs amplified diodes (Thorlabs PDA10CS). Lock-in detection was used to improve the signal-to-noise ratio.  Wavelength scanning and data acquisition were controlled by a PC.

\begin{figure}[ht]
\begin{center}
\includegraphics[width=\columnwidth]{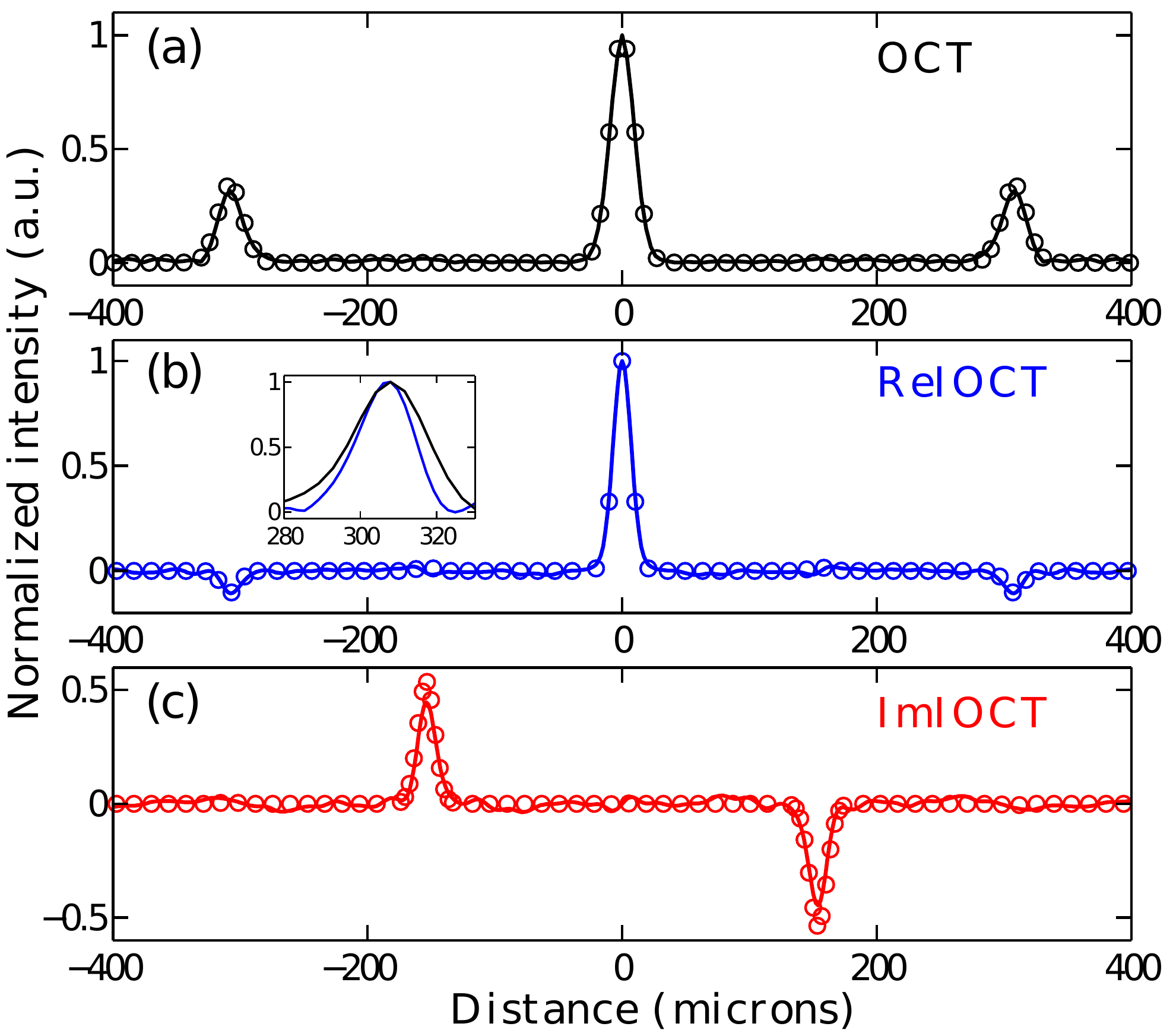}
\end{center}
\caption{(color online) Comparison of conventional OCT and intensity-based SIOCT in the absence of dispersion in the system.  The solid lines show the experimental results and the circles correspond to theoretical predictions. (a) OCT result. (b) and (c) Real and imaginary part of the Fourier transform of the SIOCT interferogram, respectively. The inset in (b) displays a direct comparison of the axial resolution of both techniques: black OCT, blue SIOCT. The intensity units have been normalized in this case.}
\label{fig:nodisp}
\end{figure}

In order to confirm the resolution improvement we first performed measurements in the absence of dispersion in the sample arm.  The results are shown in Fig.~\ref{fig:nodisp} where both the real and the imaginary part of the Fourier transform of the cross-product function $S(\omega^\prime)$ are presented. For comparison, the theoretical results obtained on the basis of Eqs.~\eqref{OCTa_sig} and \eqref{IOCT_sig} are superimposed as circles. The optical path delay was converted into physical distance as light would travel in vacuum and thereby the $0$ point represents equal path lengths. For ease of comparison with the conventional OCT result, the distance was divided by 2 for the SIOCT measurement so that the image peak would correspond to the actual sample position. In general we observe excellent agreement between the experimental and theoretically predicted results. Specifically, we see that the measured imaginary part of the Fourier spectrum is an odd function and corresponds precisely to the Fourier transform of the term $c_1(\omega^\prime)$, while the measured real part is even and matches closely the Fourier transforms of the terms $c_0(\omega^\prime)$ and $c_2(\omega^\prime)$ of the SIOCT interferogram. Significantly, we also observe how the resolution is improved in comparison to the standard OCT measurement (see inset in Fig.~\ref{fig:nodisp}). The FWHMs of the OCT and SIOCT peaks corresponding to the sample position, summarized in Table~\ref{table:nonlin}, are 23.5~$\mu$m and 17.3~$\mu$m, respectively, which gives a ratio of 0.736, close to the $\sqrt{2}$ theoretical resolution improvement predicted for a light source with a Gaussian spectrum.

\begin{figure}[t]
\begin{center}
\includegraphics[width=\columnwidth]{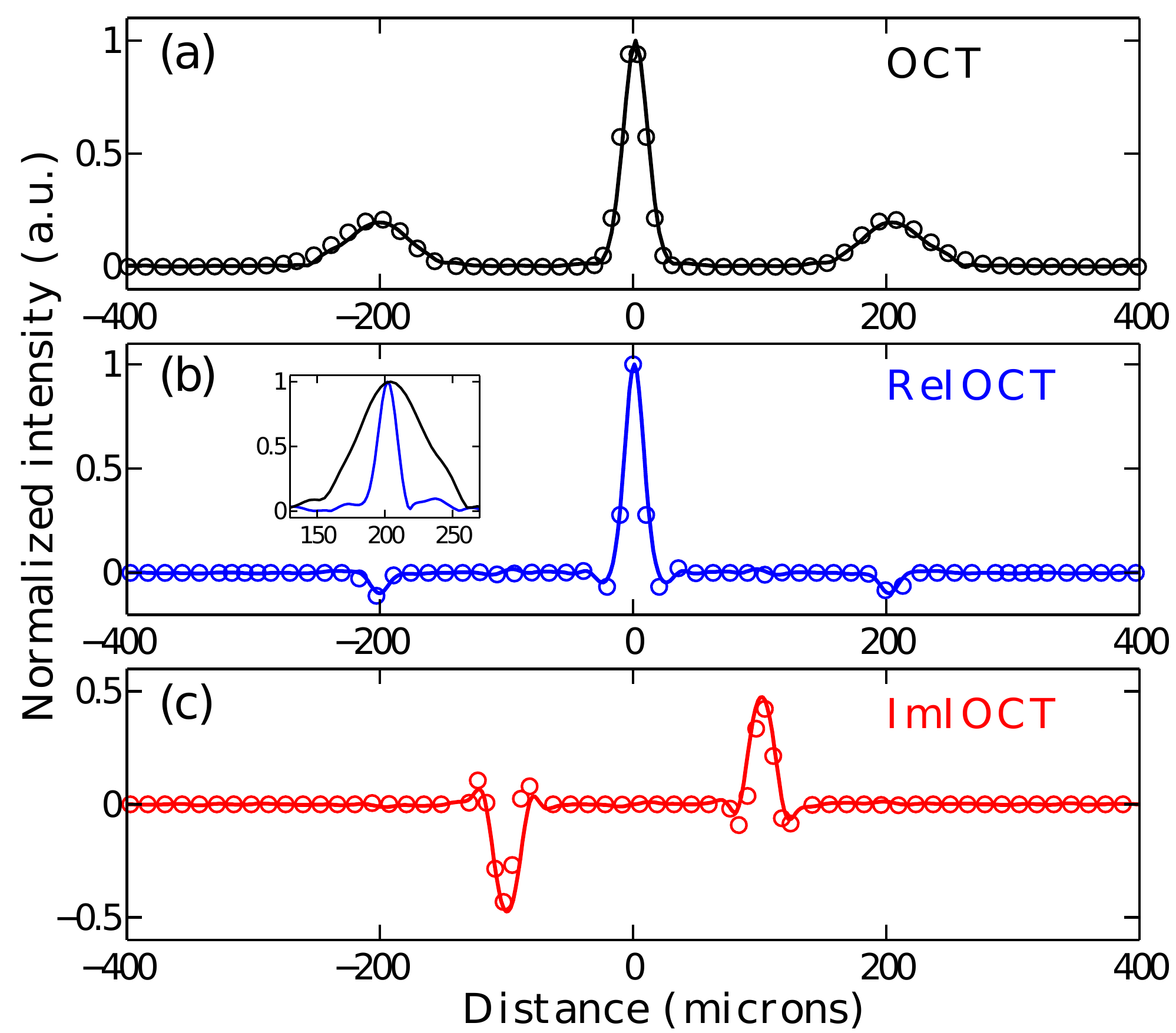}
\end{center}
\caption{(color online) Comparison of conventional OCT and intensity-based SIOCT when dispersion is present in the system.  The solid lines show the experimental results and the circles give the theoretical predictions that are based on Eqs.~\eqref{OCTa_sig} and \eqref{IOCT_sig}. (a) OCT result. (b) and (c) Real and imaginary part of the Fourier transform of the SIOCT interferogram, respectively.  The inset in (b) illustrates the direct comparison of the axial resolution in the two techniques: black OCT, blue SIOCT. The intensity units have been normalized in this case.}
\label{fig:disp}
\end{figure}

We next proceeded to confirm experimentally the inherent dispersion cancellation of the SIOCT scheme. For this purpose, an 8 cm thick bulk piece of SF10 glass was inserted into the sample arm. In order to increase the effect of dispersion the beam was made to propagate 8 times through the glass cube, effectively corresponding to a 64 cm glass piece. The group-velocity dispersion coefficient of SF10 at 1610 nm is $\beta_2 = 1.634 10^{-26} \rm{s^2/m}$, which should lead to a resolution decrease by a factor of about 2.4 in the case of conventional OCT. The measurement results for conventional OCT and SIOCT are illustrated in Fig.~\ref{fig:disp}. As for the dispersionless case, the theoretical results computed from Eqs.~\eqref{OCTa_sig} and \eqref{IOCT_sig} are also superimposed as circles. It can readily be seen that the experimental results in both cases follow the theoretical predictions with great accuracy. More specifically, we observe how in the case of conventional OCT the FWHM of the Fourier transform peak corresponding to the sample location has broadened nearly 2.5 times to 58.0~$\mu$m due to the dispersion of the SF10 glass, a value very close to that expected from the theoretical calculations. Most remarkably, the width of the SIOCT Fourier transform peak representing the sample position is unaffected by the presence of the bulk piece of glass, clearly demonstrating that dispersion is inherently canceled in SIOCT. In fact, the resolution was found to be 16.2~$\mu$m, which is even slightly better than in the absence of dispersion. This slight change in the resolution is caused by third-order dispersion, which is not canceled in SIOCT. The measured FWHMs, 16.2~$\mu$m and 58.0~$\mu$m, of respectively the OCT and SIOCT peaks corresponding to the sample position are summarized in Table~\ref{table:nonlin}, together with the dispersion-free case. Finally, we also see how the artefact peak corresponding to the term $c_1(\omega^\prime)$ is generally affected by dispersion, which provides an additional means of discrimination.

In the case of multiple interfaces, however, the artifacts have generally complex values due to cross-talk between the terms corresponding to the multiple interferences. This means that they can not be discriminated from the real interfaces by solely considering the real part of the Fourier transform of the SIOCT interferogram. However, in this case, one can take advantage of the fact that the amplitude of the artifacts corresponding to the $c_1(\omega^\prime)$ term in Eq.~(\ref{eq:c1}) depends on the center frequency $\omega_0$ of the source spectrum. Indeed, for a stationary light source the spectrum need not to be symmetrical with respect to $\omega_0$, so that one can choose arbitrarily the center frequency as long as the detected signal significantly exceeds the measurement noise. Because the phase of the $c_1(\omega^\prime)$ term oscillates with $\omega_0$, potential artifacts will then vanish when multiple measurements performed with spectra of different center frequencies are averaged. Interestingly, multiple measurements with a different center frequency can be numerically performed by post-selection of a smaller number of points from the measured interferogram, which in turn shifts the spectral window together with the central frequency. In this way, one can produce an ensemble of data sets corresponding to spectra with different central frequencies and averaging the Fourier spectra eliminates any artifact due to the presence of multiple interfaces. Yet, it is important to emphasize that averaging has its limitations as the dispersion coefficients $\beta_i$ depend on the frequency $\omega_0$ around which the Taylor expansion is performed. This leads to the situation where different central frequencies give different sample positions and, in that case, averaging may lead to broadening of the image peak and a decrease in the overall resolution.

\begin{table}[h]
\caption{Summary of resolution comparison} 
\centering 
\begin{tabular}{l c c} 
\hline\hline \\ [-1.5ex]  
   & SIOCT & Standard OCT   \\ [0.5ex] 
\hline \\ [-.5ex]
Without dispersion & 17.3 $\mu$m & 23.5 $\mu$ m \\ 
With dispersion & 16.2 $\mu$m & 58.0 $\mu$m \\ [1ex] 
\hline 
\end{tabular}
\label{table:nonlin} 
\end{table}

In summary, we have experimentally demonstrated spectral intensity optical coherence tomography and proved the associated resolution improvement and dispersion cancellation. We also suggested a possible approach to eliminate artifacts that might be present in the case of samples with multiple distinct interfaces. No fast detectors are needed in our technique as only mean spectral intensities are measured. The method works with classical light sources and standard photodiodes, illustrating the simplicity of the technique and showing its potential application to high resolution imaging.

We are grateful to Tomohiro Shirai for useful discussions. We acknowledge the Academy of Finland for financial support (project 268480).



\end{document}